\documentclass[a4paper, amsfonts, amssymb, amsmath, reprint, showkeys, nofootinbib, twoside]{revtex4-1}
\usepackage[english]{babel}
\usepackage[utf8]{inputenc}
\usepackage[colorinlistoftodos, color=green!40, prependcaption]{todonotes}
\usepackage{amsthm}
\usepackage{mathtools}
\usepackage{physics}
\usepackage{xcolor}
\usepackage{graphicx}
\usepackage[left=23mm,right=13mm,top=35mm,columnsep=15pt]{geometry} 
\usepackage{adjustbox}
\usepackage{placeins}
\usepackage[T1]{fontenc}
\usepackage{lipsum}
\usepackage{csquotes}
\usepackage[pdftex, pdftitle={Article}, pdfauthor={Author}]{hyperref}
\begin{document}
\title{Ultra-stable and versatile high-energy resolution setup for attosecond photoelectron spectroscopy}

\author{Sizuo Luo$^{1,*}$, Robin Weissenbilder$^{1,*}$, Hugo Laurell$^{1}$, Mattias Ammitzb\"{o}ll$^{1}$, Vénus Poulain$^{1}$, David Busto$^{1,2}$, Lana Neori\v{c}i\'{c}$^{1}$, Chen Guo$^{1}$, Shiyang Zhong$^{1}$, David Kroon$^{1}$, Richard J Squibb$^{3}$, Raimund Feifel$^{3}$,  Mathieu Gisselbrecht$^{1}$, Anne L'Huillier$^{1}$ and Cord L Arnold$^{1}$}

    \affiliation{$^{1}$Department of Physics,  Lund University,  Box 118,  22100 Lund, Sweden} 
      \affiliation{$^{2}$Physikalisches Institut, Albert-Ludwigs-Universit\"at Freiburg, Hermann-Herder-Stra{\ss}e 3, 79104 Freiburg,
Germany} 
\affiliation{$^{3}$Department of Physics, University of Gothenburg, Origov\"agen 6B, SE-41296 Gothenburg, Sweden}

    \email[Correspondence email address: ]{sizuo.luo@fysik.lth.se, robin.weissenbilder@fysik.lth.se}

\date{\today} 

\begin{abstract}
Attosecond photoelectron spectroscopy is often performed with interferometric experimental setups that require outstanding stability.
We demonstrate and characterize in detail an actively stabilized, versatile, high spectral resolution attosecond beamline. 
The active-stabilization system can remain ultra-stable for several hours with an RMS stability of 13\,as and a total pump-probe delay scanning range of $\sim400$\,fs.
A tunable femtosecond laser source to drive high-order harmonic generation allows for precisely addressing atomic and molecular resonances. 
Furthermore, the interferometer includes a spectral shaper in 4f-geometry in the probe arm as well as a tunable bandpass filter in the pump arm, which offer additional high flexibility in terms of tunability as well as narrowband or polychromatic probe pulses.
We show that spectral phase measurements of photoelectron wavepackets with the rainbow RABBIT technique (reconstruction of attosecond beating by two photon transitions) with narrowband probe pulses can significantly improve the photoelectron energy resolution. 
In this setup, the temporal-spectral resolution of photoelectron spectroscopy can reach a new level of accuracy and precision.
\end{abstract}


\maketitle
\section{INTRODUCTION}

The motion of electron wavepackets in small quantum systems occurs on the attosecond time scale, and in the past two decades there has been a significant effort made to track these electron dynamics in real-time \cite{Krausz,MauroNisoli}. The discovery of high-order harmonic generation (HHG) in gases \cite{McPherson, Ferray} at the end of the 80's opened the door to the production of isolated attosecond pulses and attosecond pulse trains, which were experimentally demonstrated at the beginning of this millennium \cite{Paul, Mairesse, Hentschel, Goulielmakis, Zhao}. The development of attosecond technology, i.e. producing extreme ultraviolet (XUV) attosecond pulses, and controlling the delay between these pump pulses and infrared (IR) probe pulses with attosecond precision, has opened up the world of attosecond science, and in particular made possible monitoring of electron dynamics. 
Ionization time delays in atomic  \cite{Schultze, klunder} or molecular \cite{Huppert_PRL_2016,Vos_Science2018} systems as well as from surfaces \cite{CavalieriNature2007}, the temporal buildup of electron wavepackets from an autoionizing state \cite{Kotur, Gruson} and the electronic rearrangement following  ionization have been measured on the attosecond time scale \cite{Drescher, Mansson, Zhong2020}.
Correlations between electrons during shake-up ionization \cite{Ossiander}, the interplay between electronic and nuclear degrees of freedom in molecules \cite{Cattaneo, Vos}, as well as the dynamics of shape resonances in various systems have also been explored \cite{Huppert, Nandi, Gong, Zhong2020}. 

All these attosecond experimental measurements require ultra-stable beamlines for synchronizing the pump and probe pulses to within a few attoseconds, requiring that the optical path lengths need to be stabilized on the nanometer scale. Different beamline configurations, collinear \cite{Zair, Ahmadi} or noncollinear \cite{Chini, Sabbar, HuppertRSI, Weber}, have been successfully used to achieve attosecond stability for several hours. Collinear configurations, using split-plates or mirrors are intrinsically stable since the pump and probe are propagating collinearly. However, the scanning range for the pump or probe delay is usually limited to a few tens of femtoseconds. Noncollinear setups, requiring beam splitting and recombination, provide more versatility. However, since the two optical paths propagate separately anywhere from tens of centimeters to tens of meters, the system is more susceptible to fluctuations and drifts. To counteract this, Mach-Zehnder interferometers with copropagating phase-locked continuous wave (CW) lasers have been designed and implemented, allowing for an active stabilization of the optical paths down to tens of attoseconds \cite{Chini, Sabbar, HuppertRSI, Weber, Vaughan, Mandal}. Other schemes have also demonstrated good stability, e.g. with a femtosecond IR laser stabilization system \cite{Schlaepfer, Osolodkov, Srinivas} or by stabilizing the delay with the sideband signal from a reconstruction of attosecond beating by interference of two-photon transition (RABBIT) measurements \cite{Luttmann}.

Many experiments in attosecond science require not only temporal but also spectral resolution \cite{Isinger2}. This is the case for studying the dynamics of discrete or quasi-discrete (Fano) resonances in atomic systems, for resolving vibrational structure in molecular photoionization \cite{Nandi, Haesler, Wang} or when multiple processes, e.g. due to shake-up, create spectral congestion \cite{Drescher, Mansson, Zhong2020,Kotur,Gruson,Busto2018,BustoEPJD2022}. This is especially important for vibration-resolved molecular ionization of molecules. A high spectral resolution can be obtained in photoelectron spectroscopy with highly-resolving spectrometers as well as with narrow bandwidth probe fields \cite{BustoEPJD2022}. Other factors may also affect the temporal stability and/or spectral resolution, such as the frequency chirp of the high-order harmonics or the temporal (spectral) properties of the probe field. In addition, any spatial inhomogeneity, regarding for example, the delay, will reduce the accuracy \cite{Isinger1}.

\begin{figure*}[!t]
\centering
\includegraphics[width=0.95\textwidth]{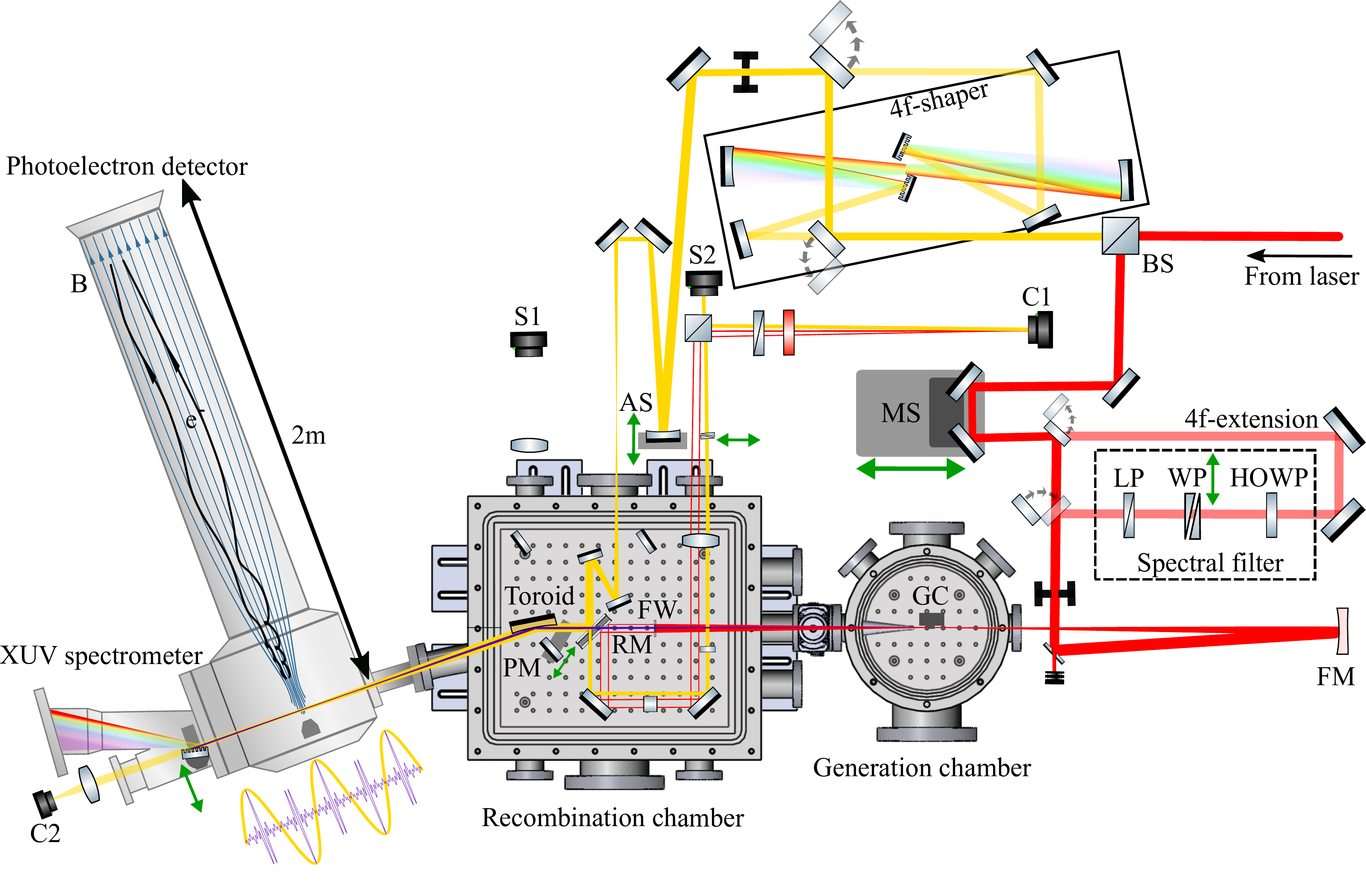}
\caption{Schematic representation of the combined attosecond beamline and MBES apparatus. The setup consists of eight main parts: a Mach-Zehnder interferometer, a femtosecond laser stabilization system, a high-order harmonic generation chamber, a recombination chamber, a 2\,m long MBES, an XUV spectrometer, a 4f-shaper, and a spectral filter. The probe beam of the interferometer is shown in yellow, the pump beam in red, and the XUV in purple. BS, beam splitter; MS, femtosecond scale manual delay stage; AS, attosecond scale piezo stage; FM, focusing mirror; GC, gas cell; FW, filter wheel; RM, recombination mirror; PM, pick-up mirror; LP, linear polarizer; WP, wedge pair; HOWP, high-order waveplate; C1, C2, cameras; S1, S2, IR spectrometers.}
\label{figSetup1}
\end{figure*}

In this work, we present an improved experimental setup for interferometric photoelectron measurements, such as RABBIT \cite{Paul,Mairesse}, with long-term stability, and high temporal and spectral resolution. Femtosecond laser pulses are split in a Mach-Zehnder interferometer, which is partly in vacuum and partly in air, thus allowing for flexibility. The delay between the pump and probe pulses is stabilized to around 13\,as root-mean-square (RMS) error over hours. A high spectral resolution in the energy-resolved (``rainbow'') RABBIT technique \cite{Gruson} is obtained by using a magnetic bottle electron spectrometer with a 2\,m long flight tube and a narrowband wavelength filter in the probe arm \cite{BustoEPJD2022}. We demonstrate the long-term stability of our interferometer by performing a scan with a 70\,fs delay variation over seven hours, and we show the improved spectral resolution due to the narrowband probe (10\,nm bandwidth around 790\,nm) in a RABBIT measurement across the spin-orbit resolved 3s$^{-1}$4p resonance in argon. Our experimental setup includes a 4f-shaper, which can be coupled in and out of the probe beam path, which allows shaping the bandwidth of the probe laser into mono-, bi- or polychromatic narrowband components, easily tunable within the laser bandwidth. The 4f-shaper allows the implementation of new, advanced, methods for the characterization of photoelectron wavepackets, using, for example, quantum state tomography \cite{Laurell}. 

\section{EXPERIMENTAL SETUP}

\subsection{Overview}

\begin{figure}
\begin{center}
\includegraphics[width=0.45\textwidth]{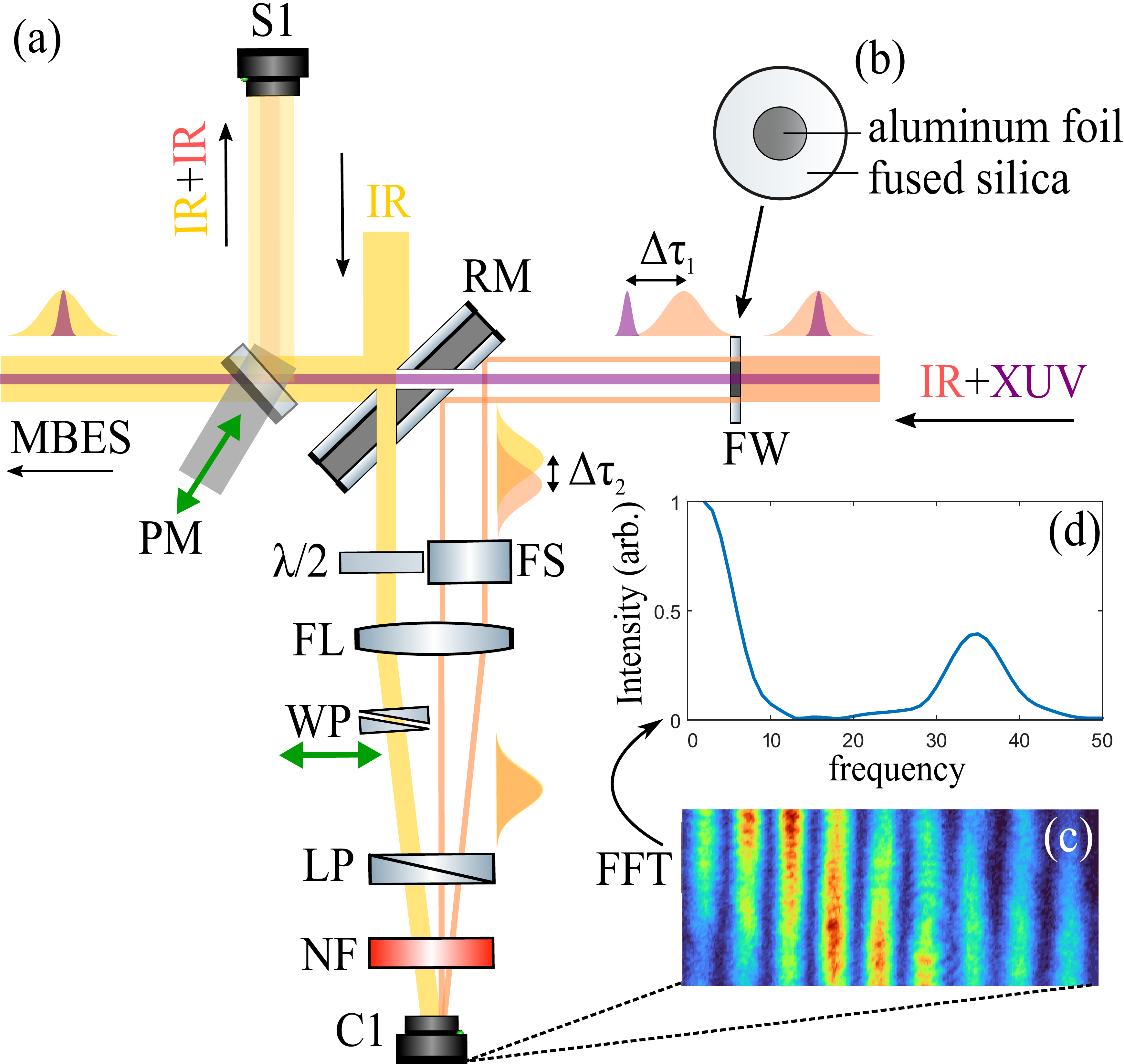}
\end{center}
\caption{(a) An illustration of the essential components for measuring the temporal stability of the interferometer, with the probe beam (yellow lines), pump beam (red lines) and XUV (purple lines). FW, filter wheel; RM, recombination mirror; PM, pick-up mirror; $\lambda/2$, half-waveplate; FS, fused silica; FL, focusing lens; WP, wedge pair; LP, linear polarizer; NF, narrow bandpass filter; MBES, magnetic bottle electron spectrometer. (b) schematic of the aluminum foil mounted on fused silica. (c) Spatial fringes from interfering the femtosecond lasers (d) Fourier transform of the fringe pattern.}
\label{figSetup2}
\end{figure}

A schematic drawing of the experimental setup is shown in Fig.~\ref{figSetup1}. It consists of the pump (red lines), probe (yellow lines) and XUV (purple lines) beam paths, the stabilization system, an HHG chamber, a recombination chamber, and a magnetic bottle electron spectrometer (MBES).
Briefly, the beam from a Titanium-Sapphire chirped pulse amplification laser system (1-3\,kHz repetition rate, up to 4\,mJ pulse energy, tunable 40-80 nm bandwidth and tunable central wavelength around 800\,nm) is split by a broadband beam-splitter (BS) into the pump (40\%) and probe (60\%) arms.
The beam-splitter is the first optical element in the Mach-Zehnder interferometer and active stabilization system, details of which are presented in the next section. 

In the pump arm, the beam can pass through an optional spectral filter (discussed below) or directly be apertured by a holey mirror, removing the central part of the beam.
The annular beam is focused by a dielectric concave mirror (FM, focal length $f=50$\,cm) into a gas cell (GC) where high-order harmonics are generated. The cell has a length of 6\,mm and is mounted on a stage which allows for fine tuning of its position along five dimensions ($x$, $y$, $z$, $\theta_y$, $\theta_z$), for alignment relative to the laser and also for the control of the phase matching conditions. The gas is supplied by a  pulsed piezo-valve (Amsterdam Piezo Valve) with variable opening time to control the pressure in the cell. Typical conditions imply a backing pressure of around 4--6\,bar, resulting in an ambient pressure of tens of mbar in the gas cell. The collinearly-propagating IR and generated XUV beams pass on a holey fused silica plate, mounted on a rotatable filter wheel (FW). The 1\,mm thick fused silica plate with a 2\,mm diameter hole, covered by an aluminum foil of 200\,nm thickness, blocks the inner part and transmits the outer part of the pump IR. The aluminum foil transmits the XUV beam which then passes through the holey recombination mirror (RM). The toroidal mirror performs the 1-to-1 imaging of the XUV beam from the generation into the gas jet of the MBES. The outer part of the pump IR is reflected by the RM and is used for the temporal stabilization of the system. The RM is drilled in two directions to transmit and reflect parts of both the pump- and probe arms of the interferometer, which is integral to the active delay stabilization. The stabilization will be discussed in detail in section \ref{sec:stabilization}. 

The probe IR laser either passes through a 4f-shaper (discussed below) or is directly focused into the recombination chamber by a silver coated concave mirror ($f = 50$\,cm) mounted on a piezo delay stage (AS) for the purpose of controlling the delay between the two pulses with attosecond scale precision. The outer part of the probe beam is reflected by the RM and focused into the MBES with the toroidal mirror, where it is spatio-temporally overlapped with the XUV in the experiment. The inner part of the probe beam passes through the hole in the RM, to be used for stabilization, achieving attosecond temporal resolution. There is a long--delay stage (MS) in the beam path of the pump for coarse adjustments of the delay between the two arms on the femtosecond scale, while the piezo delay stage in the probe arm is used to control the delay from tens of attoseconds to hundreds of femtoseconds. 

The MBES has a 2\,m long flight tube with a $4\pi$\,sr collection angle. The energy resolution is better than 80 meV for low energy photoelectrons and retarding lenses can slow down fast electrons to enhance the resolution of high energy electrons. After the interaction chamber, the XUV and IR beams reach an XUV spectrometer chamber where the harmonic spectrum and divergence, or the IR spatial profile, can be monitored. The home-made flat-field XUV spectrometer is based on a Hitachi concave grating, which images the focus in the interaction chamber onto the plane of the detector, consisting of multichannel plates and a phosphor screen.
The grating is placed on a linear translation stage so that it can be moved in or out of the beam path. Moving the grating out and rotating the aluminum filter out of the pump arm, the spatial overlap of the two arms in the interaction can be imaged at the camera (C2). When coupling the probe into the 4f-shaper, a polychromatic narrow bandwidth spectrum can be obtained by placing slits in the Fourier plane of the shaper. The shaper consists of two silver coated concave mirrors ($f = 50$\,cm) and two gratings with 1200 lines per mm, purchased from Spectrogon. The increased beam path length due to the 4f-shaper is compensated by extending the pump arm by the same distance (4f-extension), hence preserving the temporal overlap and, to a large extent, stability. 

\subsection{Active stabilization}
\label{sec:stabilization}

\begin{figure*}[!t]
\begin{center}
\includegraphics[width=1\textwidth]{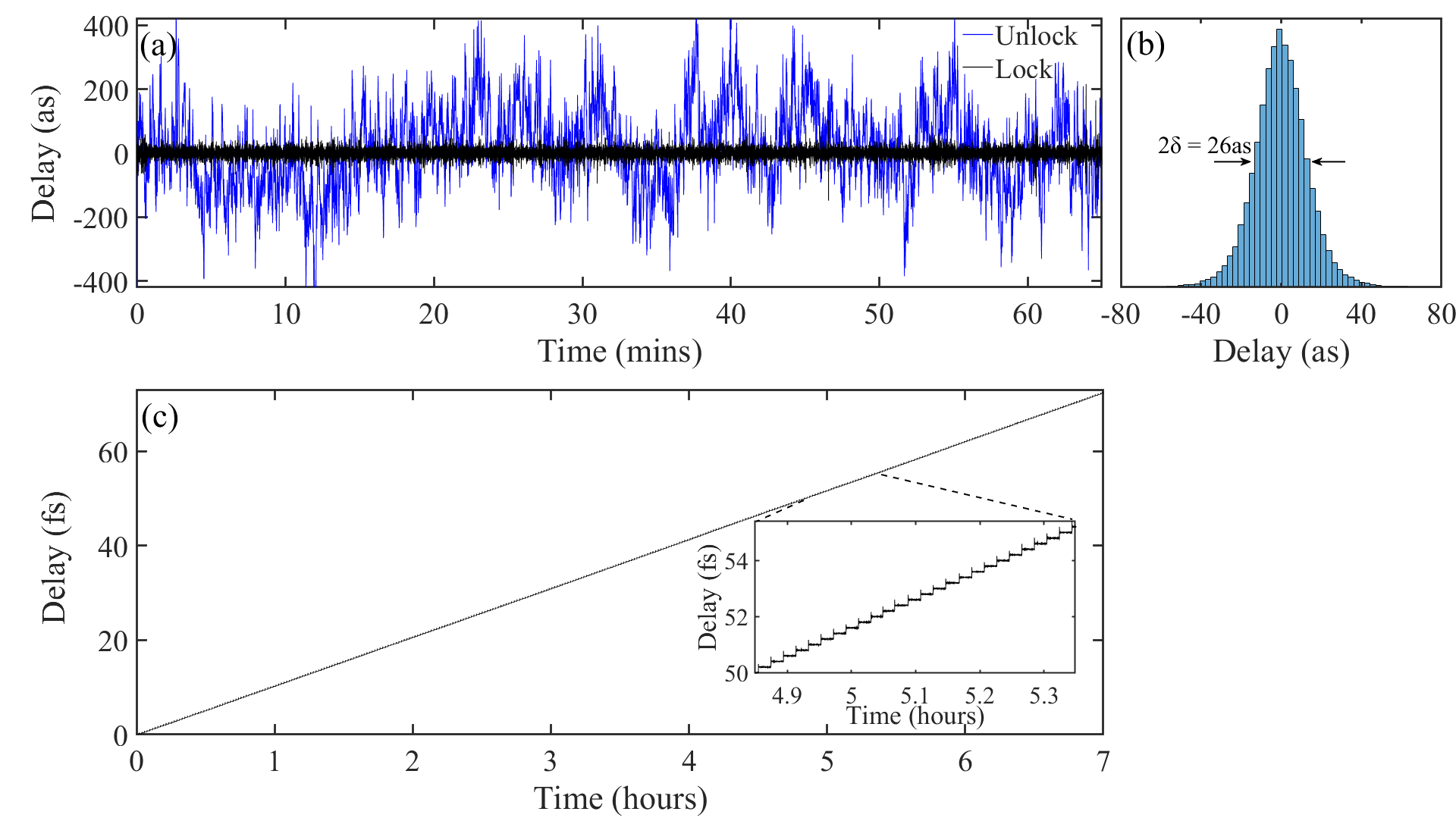}
\end{center}
\caption{(a) Variation of the long-term relative delay over time when the interferometer is unlocked and locked, (b) Standard deviation of the delay when the stabilization is active over one hour. The RMS is around 13\,as. (c) Variation of the delay during a long scan; the inset shows the effect of the delay control.}\label{figLongstab}
\end{figure*}

To stabilize the pump-probe setup on the attosecond time scale, we derive an error signal from the interference of parts of the IR from the pump- and probe arms of the interferometer (AS in Fig.~\ref{figSetup1}).
The layout of the stabilization system is shown in Fig.~\ref{figSetup2}(a). The outer part of the pump laser (red lines), passing through the fused silica filter plate [shown in Fig.~\ref{figSetup2}(b)], is reflected by the RM, whereas the inner part of the probe laser (yellow line) propagates through it. The two beams are transversely shifted after the RM, and when focused by a lens (FL) and overlapped in time, give rise to spatial fringes detected by a camera, as shown in Fig.~\ref{figSetup2}(c).
Fast Fourier transform (FFT) is used  to identify the spatial frequency component corresponding to the fringes [shown in Fig.~\ref{figSetup2}(d)] and to find their phase. 
 An error signal, generated from the difference between the desired and the measured phases, is used to stabilize the interferometer with a feedback loop controlling the piezoelectric translation stage in the probe arm (see AS in Fig.~\ref{figSetup1}). The fringes are recorded at 13\,Hz. 
A long illumination time of 60\,ms per frame, averaging 180 consecutive shots at 3\,kHz repetition rate of the laser, is chosen to compensate for the slow delay drifts during RABBIT measurements.

RABBIT measurements require that the XUV and IR pulses overlap in time in the interaction region in the MBES and at the same time the two IR pulses used for temporal stabilisation must overlap at the camera C1.
In this sense, our stabilization scheme is conceptionally different from those based on CW lasers, which neither give any information on the actual XUV and IR temporal overlap in the interaction region nor are sensitive to the gas target density used for HHG \cite{DavidKroon}.
Once the temporal overlap is achieved both in the MBES interaction region and at camera C1, it can easily be maintained, on a daily basis, by checking the spatial fringes on the camera.
A rough estimate of the temporal overlap in the interaction region of the MBES can be obtained by using a pickup mirror (PM) to reflect the generating and probe IR pulses towards a spectrometer (see also S1 in Fig.~\ref{figSetup1}), when the aluminum filter is removed, and by tuning the coarse stage in the pump arm (FS in Fig.~\ref{figSetup1}) until the beam pathes have the same length. 
Then, the difference between the XUV and probe IR pulses results only from the transmission of the XUV through the aluminum filter. The fine adjustment needed to find exact temporal overlap is done with the help of a RABBIT measurement (see below).
To make sure that the temporal overlap in the interaction region and at the stabilization camera C1 is obtained at the same time, path length differences in the stabilization part (Fig.~\ref{figSetup2}a), originating (i) from the fused silica plate that holds the metallic filter and (ii) from geometrically different beam pathes at the RM for the pump- and probe pulses, are compensated by an additional piece of glass in one arm and a pair of glass wedges in the other arm. The contrast of the fringes recorded with C1 is optimized with a $\lambda/2$-plate in one arm and a polarizer before the camera.
Finally, a narrow spectral bandpass filter is used to extend the temporal overlap region. 
The stabilization system works up to 400\,fs scanning range when using a $(10\pm 2)$\,nm filter centered at 790\,nm. This filter can be easily changed to different central wavelengths to adjust for the tunability of our laser spectrum and different bandwidths can be used to increase or decrease the scanning range.

\subsection{Long-term stability}

\begin{figure}
\begin{center}
\includegraphics[width=0.5\textwidth]{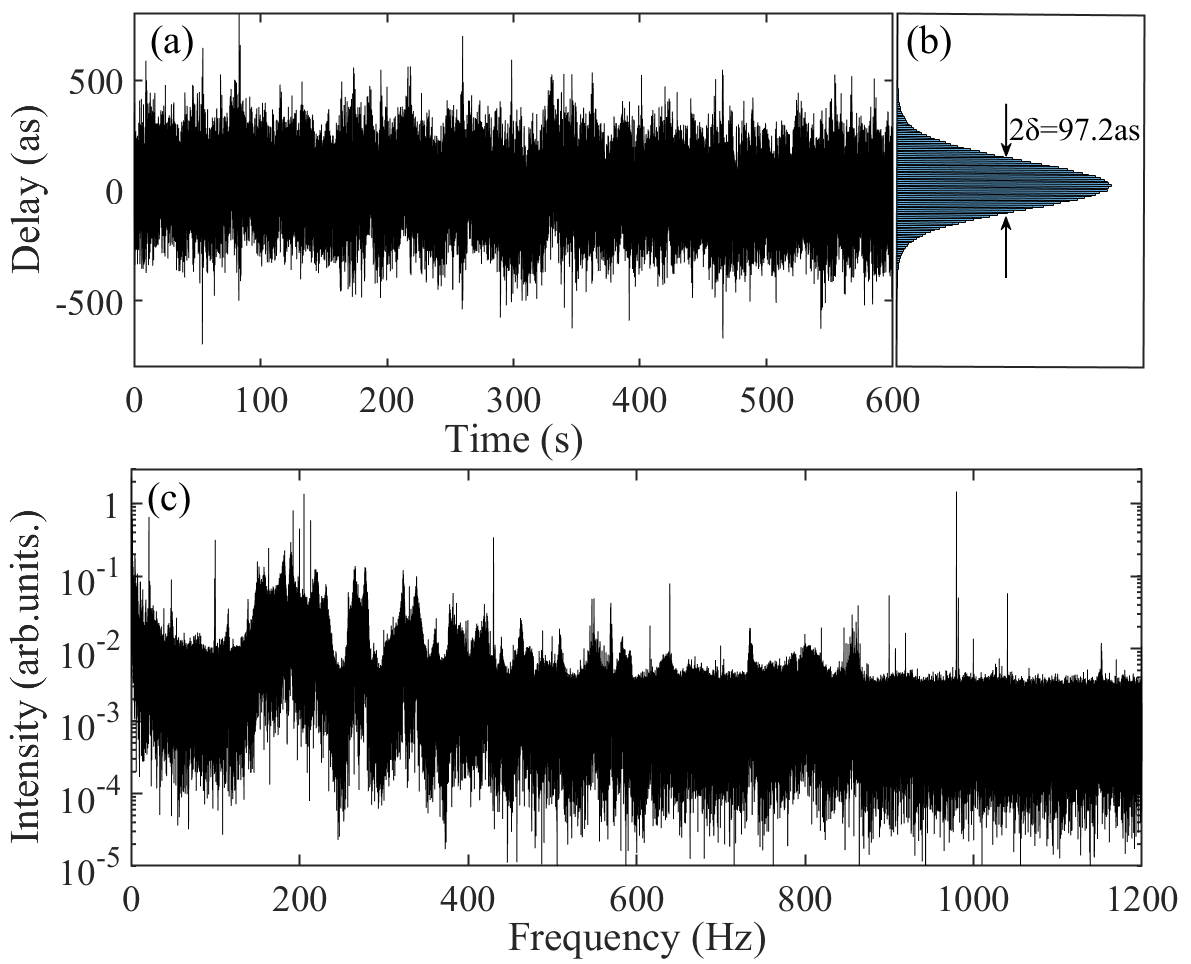}
\end{center}
\caption{(a), (b) Variation and standard deviation of the relative delay by shot-by-shot measurement at 3 kHz when the interferometer is locked. (c) Spectral composition of the short-term stability of the interferometer with active stabilization.}\label{figStab3kHz}
\end{figure}

Often, attosecond experiments require stable delay conditions over several hours, especially for high quality photoelectron spectroscopy measurements. We characterized both the short- and long-term stability of the interferometer in different conditions. In-loop measurements, i.e. the phase error extracted from the stabilization camera C1, are instructive to get an estimate of the overall stability, but do not necessarily reflect the actual stability achieved in the interaction region. The interaction region, i.e. sensitive region of the MBES, is not easily accessible for diagnostics. For out-of-loop measurements, we therefore use the PM to reflect both pump- and probe IR beams, while misaligning the RM slightly to produce spatial fringes in the detector plane. For these measurements the metallic filter was removed, including the fused silica plate carrying it. The delay shift in the stabilization beam path was adjusted with the wedge pair.
Two detectors were used, i.e. a camera at 13\,Hz frame rate, again with an illumination time of 60\,ms, for long-term measurements, and a fast CCD line detector (Basler L101k) for shot-to-shot measurements at the full 3\,kHz repetition rate of the laser.
Fig.~\ref{figLongstab}(a) shows the relative long-term time delay when the stabilization system is locked and unlocked, respectively. When unstabilized, the maximum drift of the interferometer measured during 1\,h is in the range of 400\,as. With the stabilization on, the obtained variation of the measured delay is around 13\,as RMS.
The true shot-to-shot instability, obtained with the CCD line detector, is around 50\,as with 10\,min recording time [Fig.~\ref{figStab3kHz}(a,b)]. Full repetition rate shot-to-shot measurements reveal the limiting factors for the stability via Fourier decomposition in time [Fig.~\ref{figStab3kHz}(c)].
The spectrum contains several sources of vibration in our system; most prominent are mechanical resonances between 200-400\,Hz and a vibration peak resulting from a turbomolecular pump at 980\,Hz.
A small peak at 100\,Hz originates from the Roots pumps of the central vacuum system, located in another room, on a separate, mechanically vibration-isolated, platform.
Without these precautions, the vibrations of the pumps would dominate the instability spectrum.
Finally, Fig.~\ref{figLongstab}(c) shows in-loop measurements from a delay scan taken over a total of 70\,fs in 7\,h scanning time. The stabilization system remained for 90\,s at each delay step. The inset illustrates the change of delay by moving the piezoelectric stage (AS in Fig.~\ref{figSetup1}). The measurement demonstrates the stability during long-term scans.

\label{sec:stabilizationLong}


\subsection{Spectral shaping of the probe pulse}

\begin{figure}
\begin{center}
\includegraphics[width=0.5\textwidth]{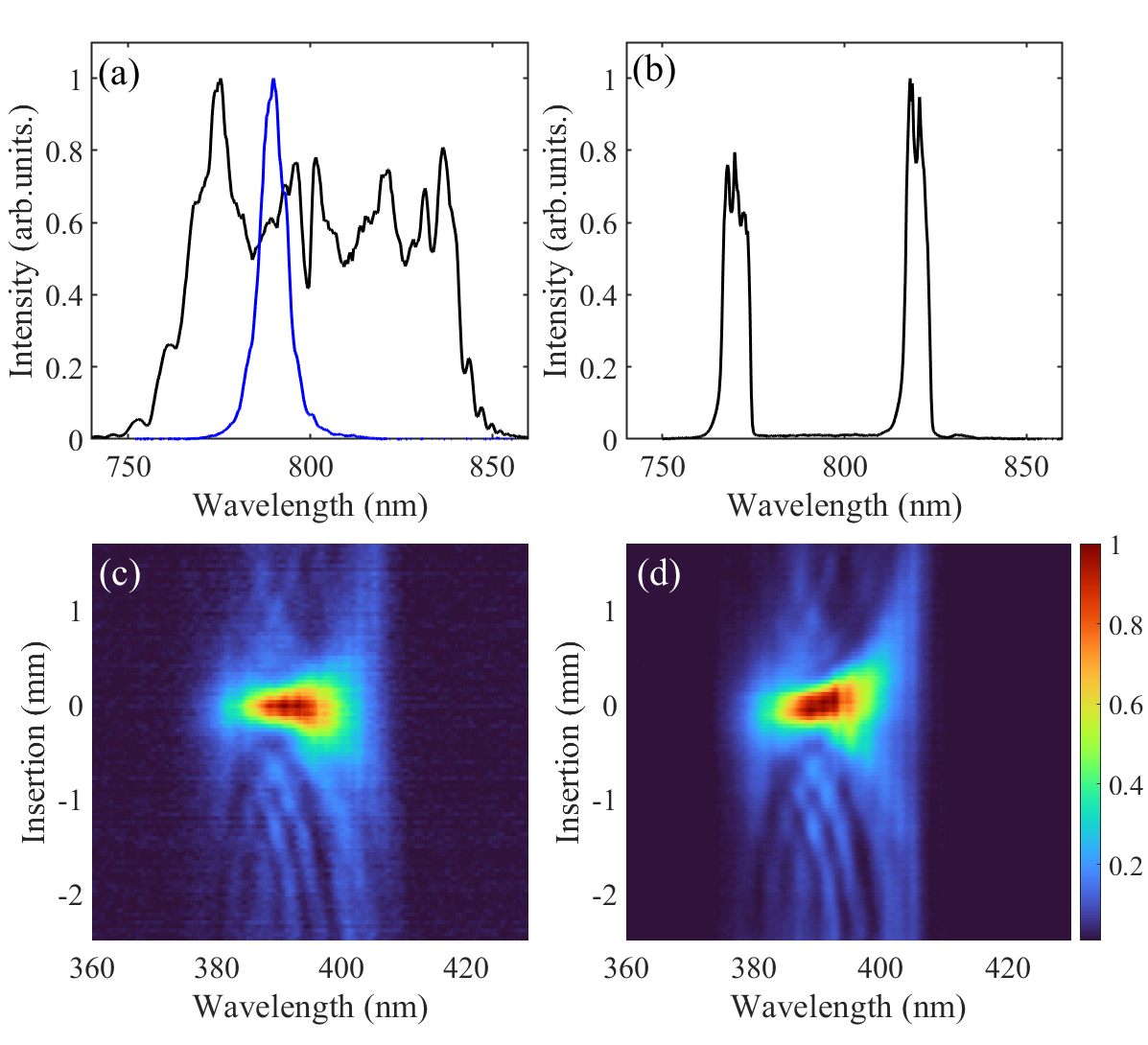}
\end{center}
\caption{(a) Spectrum from the femtosecond laser before (black) and after (blue) passing through a 10 nm narrowband filter centered at 790\,nm. (b) Spectrum after passing through the 4f-shaper with two slits in the Fourier plane. Measured d-scan trace, based on second harmonic generation, before (c) and after (d) passing through the 4f-shaper.}
\label{figIRspectra}
\end{figure}

Besides its stability, our setup offers a large flexibility for both pump and probe spectral shaping. Narrowband spectral filters can be inserted in the probe path, improving considerably the spectral resolution of RABBIT measurements, as described in the next section (see also \cite{BustoEPJD2022}). The IR spectra from our femtosecond laser before and after a 10\,nm narrow bandpass filter are compared in Fig.~\ref{figIRspectra}(a). Both the bandwidth and central wavelength of the probe can easily be changed by inserting different band pass filters.
To further improve the flexibility, a 4f-shaper was implemented into the probe arm of the interferometer, as shown in Fig.~\ref{figSetup1}.
The combination of a grating and a silver-coated concave mirror, spatially separate the spectral components in the mirror's Fourier plane, where various spectral shaping can be performed by combinations of slits of different widths.
An identical focusing mirror and grating collimate the beam and remove the angular chirp.
This technique allows creating for example bichromatic probe pulses, as shown in Fig.~\ref{figIRspectra}(b). Additionally, a spatial light modulator placed in the Fourier plane (currently not installed) can give full tunability and shaping of the probe pulses in terms of amplitude and phase.

The temporal dispersion of the 4f-setup is minimized with the help of dispersion scan (d-scan) measurements \cite{Miranda,Ivan}. Comparing the measured d-scan traces before and after the 4f-shaper, as shown in Fig.~\ref{figIRspectra}(c,d), indicates that the 4f-shaper does not increase second-order dispersion. However, some slight third-order dispersion, indicated by the tilt above 400\,nm in Fig. \ref{figIRspectra}(d), is added, which most likely is related to the large beam size and compact configuration of the shaper.

\subsection{Spectral shaping of the pump pulse}
\label{cw_pump}
The femtosecond laser chain is equipped with acousto-optic programmable gain and dispersive filters (DAZZLER and MAZZLER), which allow tuning the central wavelength and bandwidth of the laser, before being split into pump and probe.
This tunability is essential in spectroscopic applications, for example, to excite a specific resonance by absorption of a high-order harmonic with a spectrum centered on the resonance central frequency. It is however sometimes incompatible with the requirements for the probe pulse.
For e.g. the generation of a bichromatic probe pulse, a broad input spectrum to the 4f-shaper is required.
To nonetheless enable independent pump and probe spectral shaping, a compact, inline spectral filter capable of tuning the bandwidth and central wavelength was designed and placed in the pump pathway before HHG (Spectral filter in Fig.~\ref{figSetup1}).

\begin{figure}
\begin{center}
\includegraphics[width=0.5\textwidth]{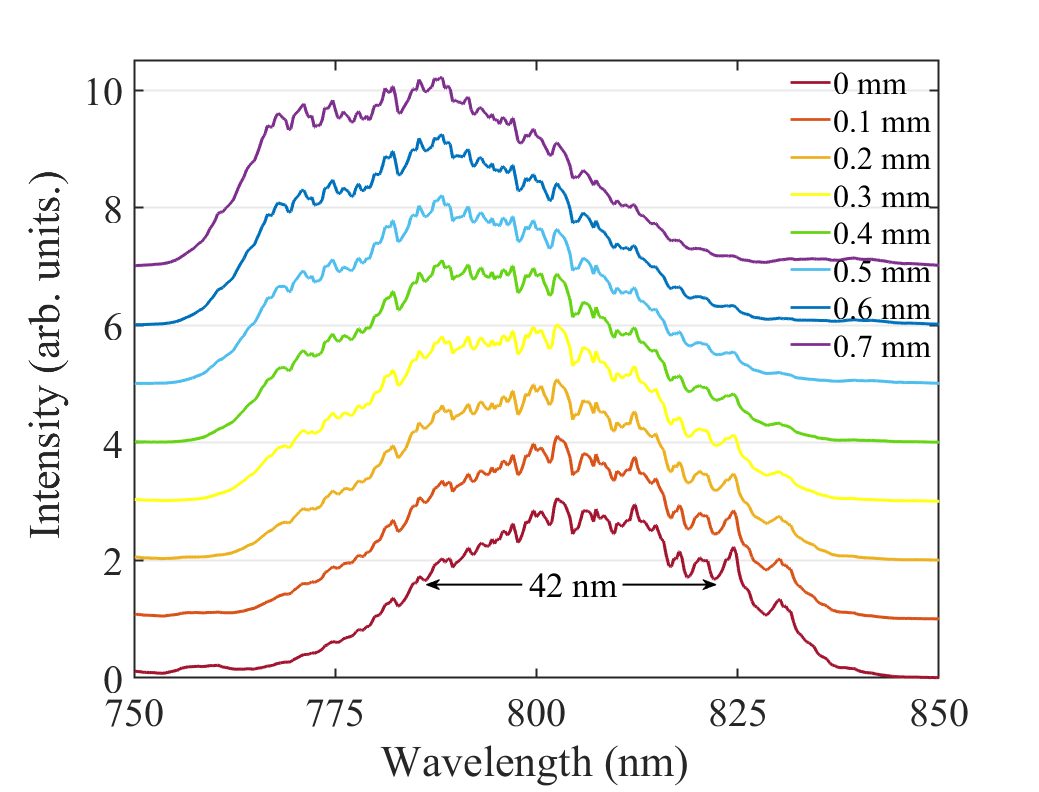}
\end{center}
\caption{Transmission spectrum of the pump pulses for different wedge positions. The thickness of the birefringent filter changes with -30\,$\mu$m per 1\,mm translation.}
\label{figIRfilter}
\end{figure}

The filter that is inspired by a single-stage Liot filter is comprised of a high-order wave plate, oriented at $45^\circ$ to the incoming linear polarization, and a polarizer behind. The inherent chromaticity of such arrangement results in a sinusoidal transmission as a function of wavelength, with the peaks, when the thickness corresponds to a high-order, full $\lambda$-plate and zeros, when the thickness corresponds to a $\lambda/2$-plate.
If the order (i.e. the thickness) of the filter is chosen carefully, mostly one lobe within a selected bandwidth is transmitted (see Fig.~\ref{figIRfilter}), where the bandwidth (FWHM) is approximately $\Delta\lambda \approx \frac{\lambda_0}{2N}$. Here, $\lambda_0$ is the selected transmission peak and $N$ is the integer order of the wave plate. 
In practice, the thickness of the wave plate must be tunable to choose the transmitted peak and bandwidth. This is achieved by a pair of anti-reflection coated quartz-wedges, cut in the $c-z$-plane, with a thickness of 0.5\,mm at the thin edge, 1.4\,mm at the thick edge and a length of 30\,mm. 
Due to the beam size in the probe arm, the minimum thickness of the plate formed by the wedges is 1.45\,mm, which corresponds to too high order ($N\approx16$ for $\lambda_0=800$\,nm). We therefore combine the wedges with a second 1\,mm thick high-order quartz wave plate, oriented with its axes crossed in respect to the first one. The combination of the two plates allows for any order between approximately 5 and 15.
For $N=10$ (i.e. $d\approx 900\,\mu$m), we obtain a transmission bandwidth of about 40\,nm  (see Fig.~\ref{figIRfilter}) and by slightly tuning the wedge position, the transmission window can be moved.

\section{EXPERIMENTAL RESULTS}

We now demonstrate the capabilities of our experimental setup with a series of experiments, testing the temporal stability and the spectral resolution, and showing the flexibility for different excitation schemes.

\subsection{Temporal stability: Long-term RABBIT experiment}

In the RABBIT experiments presented in this article, high-order harmonics were generated with a transform limited pulse of 40\,fs, centered at 800\,nm wavelength, focused into a gas cell filled with argon. After passing through the aluminum filter, the XUV radiation includes harmonics from the 11th to the 25th order. We denote these HH$q$ in the following, where $q$ is the harmonic order. Both the XUV pulse train and the probe laser are focused by the toroidal mirror into a diffusive jet of helium in the interaction region of the MBES. The pump-probe delay is varied by 70\,fs with a step size of 200\,as. 

\begin{figure}
\begin{center}
\includegraphics[width=0.5\textwidth]{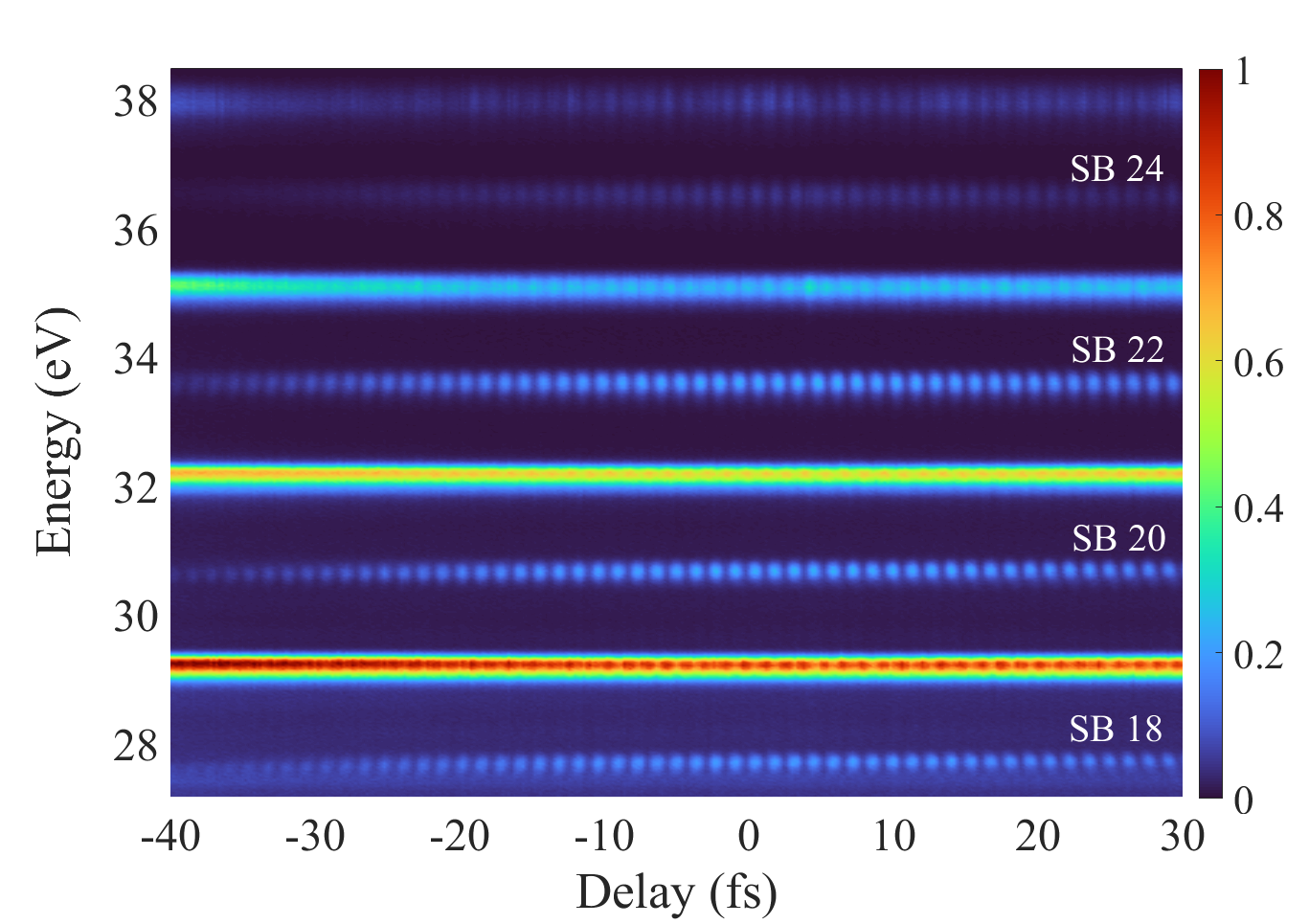}
\end{center}
\caption{Photoelectron signal (in color) in He as function of harmonic photon energy and delay in a RABBIT measurement. The delay step is 200 as.  The total acquisition time of the scan is $\sim7$ hours.}
\label{figHeRABBIT}
\end{figure}

The recorded RABBIT spectrogram is shown in Fig.~\ref{figHeRABBIT}. Besides photoelectron peaks corresponding to absorption of harmonics, sideband peaks are visible. These peaks, labeled by the net number ($n$) of involved laser photons (SB$n$), are due to two-photon transitions, absorption of HH$q$ (HH($q+2$)) and absorption (emission) of one IR photon from the probe laser. 
Due to the interference of these two pathways, the sideband signal oscillates at frequency $2\omega$, i.e. twice the angular frequency of the driving laser, with a period of $\tau_0 = 1.33$\,fs for 800\,nm \cite{Paul,Mairesse}.
The results presented in Fig.~\ref{figHeRABBIT} exhibit an excellent contrast, over a long delay range of 70\,fs, and for a total acquisition time of seven hours. This demonstrates the high stability and reliability of our interferometric setup. 

\subsection{Tunability: Below-threshold RABBIT experiment}
\begin{figure}
\begin{center}
\includegraphics[width=0.5\textwidth]{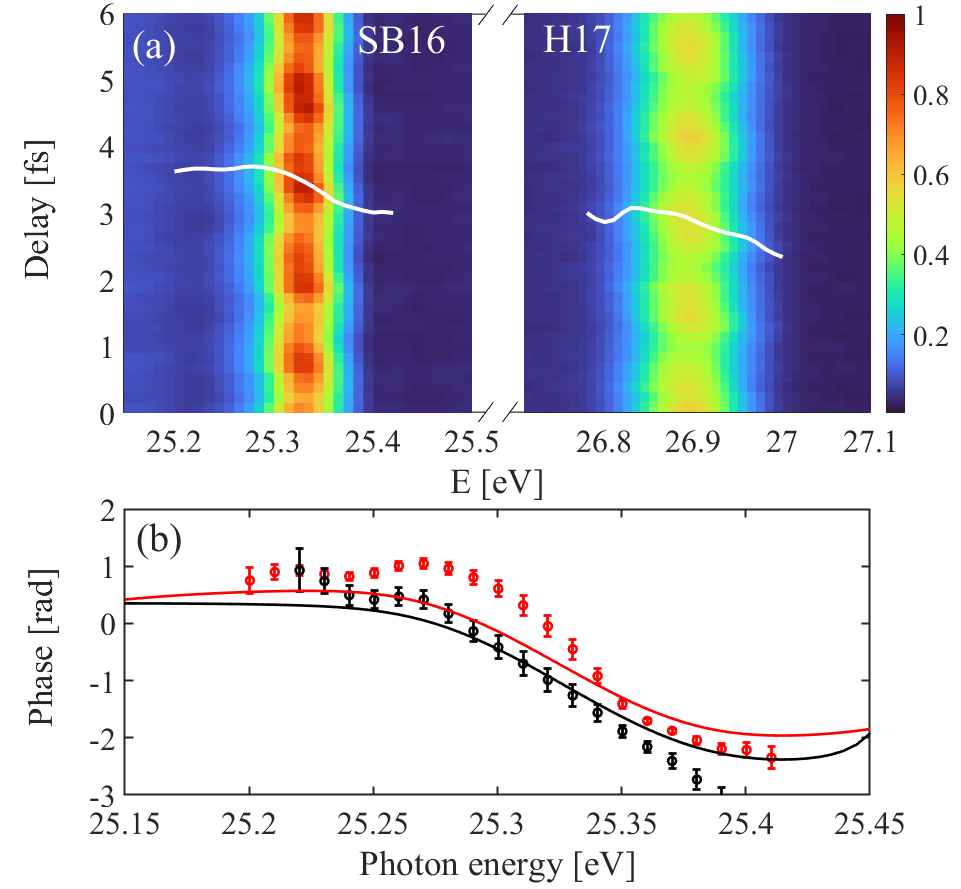}
\end{center}
\caption{(a) Photoelectron signal (in color) of SB16 and HH17 from He in a below-threshold RABBIT measurement, where HH15 excites the 1s4p state. (b) The measured phase of SB16 when the probe IR polarization is parallel (red dots) and perpendicular (black dots) to XUV. The red and black lines are calculated for IR parallel to XUV and d-wave only \cite{Lana}.}
\label{figHe4p}
\end{figure}

In below-threshold RABBIT measurements \cite{Swoboda}, one harmonic (HHq) is tuned to excite a bound state of an atom, which is ionized by absorption of an additional IR photon. The created wavepacket is probed by a nonresonant-ionization process involving absorption of HH$(q+2)$ and emission of an IR photon. This technique allows probing attosecond dynamics of resonant ionization from Rydberg states \cite{Swoboda,Kheifets,Drescher2022,autuori}. In Ref.~\cite{Lana}, we study two-photon resonant ionization of He via the 1s4p$^1$P$_1$ state which is reached by absorption of HH15, after tuning the central laser wavelength to 794 nm. The oscillation of SB16 is shown in Fig.~\ref{figHe4p}. The results show a smooth $\pi$-rad phase variation across the resonance in SB16, and a relative $\pi$-rad phase shift between SB16 and H17, as marked by the white lines in Fig.~\ref{figHe4p}(a). Furthermore, a comparison between the results obtained with the probe polarization parallel to, and perpendicular to, the polarization of the XUV light is shown in Fig.~\ref{figHe4p}(b). As explained in \cite{Lana}, by changing the polarization of the probe field, different ionization channels can be selected.  Both s and d channels coexist in the parallel conﬁguration, while only the d channel is allowed when the probe polarization is perpendicular to that of the pump since only $\Delta$m $=\pm 1$ transitions are possible. The measured relative phase shifts are well reproduced by calculations using the two-photon RPAE method \cite{vinbladh2019}.

\subsection{Spectral resolution: Resonant RABBIT experiment with a narrowband probe}
\label{2b}

\begin{figure}[!t]
\centering
\includegraphics[width=1\linewidth]{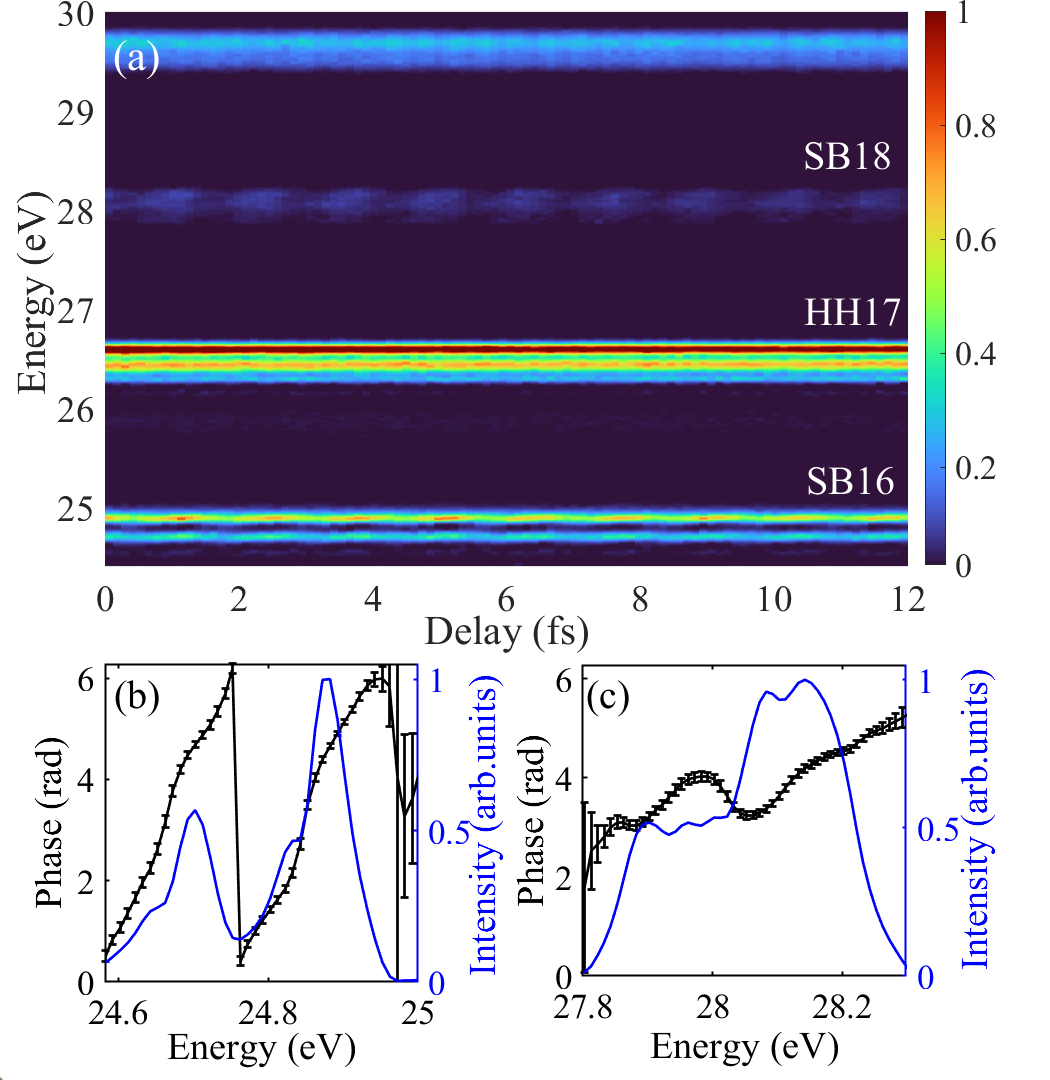}
\caption{(a) Photoelectron signal (in color) in Ar as function of harmonic photon energy and delay in a RABBIT measurement. Spectral intensity (blue) and phase (black) of the $2 \omega$ oscillation component for (b) SB16 and (c) SB18. The delay step is 120\,as. Recording this scan took $\sim3$ hours. Absorption of HH17 is resonant with the 3s$^{-1}$4p autoionizing state in Ar.}
\label{figRainbow}
\end{figure}

Fig.~\ref{figRainbow} shows a RABBIT spectrogram obtained in Ar. The central wavelength and bandwidth of the femtosecond laser is tuned (before the interferometer) so that the energy of HH17 is close to that of the autoionizing state 3s$^{-1}$4p (E$_\textrm{res} = 26.61$\,eV). A 10\,nm bandpass interference filter centered at 790\,nm [see Fig.~\ref{figIRspectra}(a)] is placed in the probe arm. A deconvolution algorithm is used to account for the spectrometer resolution \cite{Busto2018}. The photoelectron peaks present two components, due to the spin-orbit interaction, which leads to two final ion states (3p$^5\ ^2$P$_{1/2}$ and 3p$^5\ ^2$P$_{3/2}$) separated by 0.18\,eV. Both spin-orbit splitting and autoionization resonance, with the window-like spectral profiles, are visible in HH17 as well as in SB16 and SB18.  

The obtained spectral resolution allows us to perform a RABBIT analysis as a function of electron energy, a method called rainbow RABBIT \cite{Gruson}. The oscillations of the sidebands are analyzed by Fourier transform. The spectral amplitude and phase of the Fourier components oscillating at frequency $2\omega$ are shown in Fig.~\ref{figRainbow}(b,c) for SB16 and SB18. Both plots present two separate and similar components which are due to the spin-orbit splitting of the Ar$^+$ ground state, and separated in energy, as expected, by 0.18\,eV. Our energy resolution is better than in previous data, where the two spin-orbit components partially overlapped in energy \cite{Turconi}. Furthermore, in the case of SB16, the phase varies by close to $2\pi$\,rad across the Fano resonance, which is larger than previously measured values \cite{Kotur,Turconi}, thus indicating improved spectral resolution and measurement accuracy. An explanation of the $2\pi$\,rad phase jump will be presented in a future publication.

\subsection{Flexibility: KRAKEN experiment with a bichromatic probe}
\label{sec:KRAKEN}

\begin{figure}[!t]
\centering
\includegraphics[width=1\linewidth]{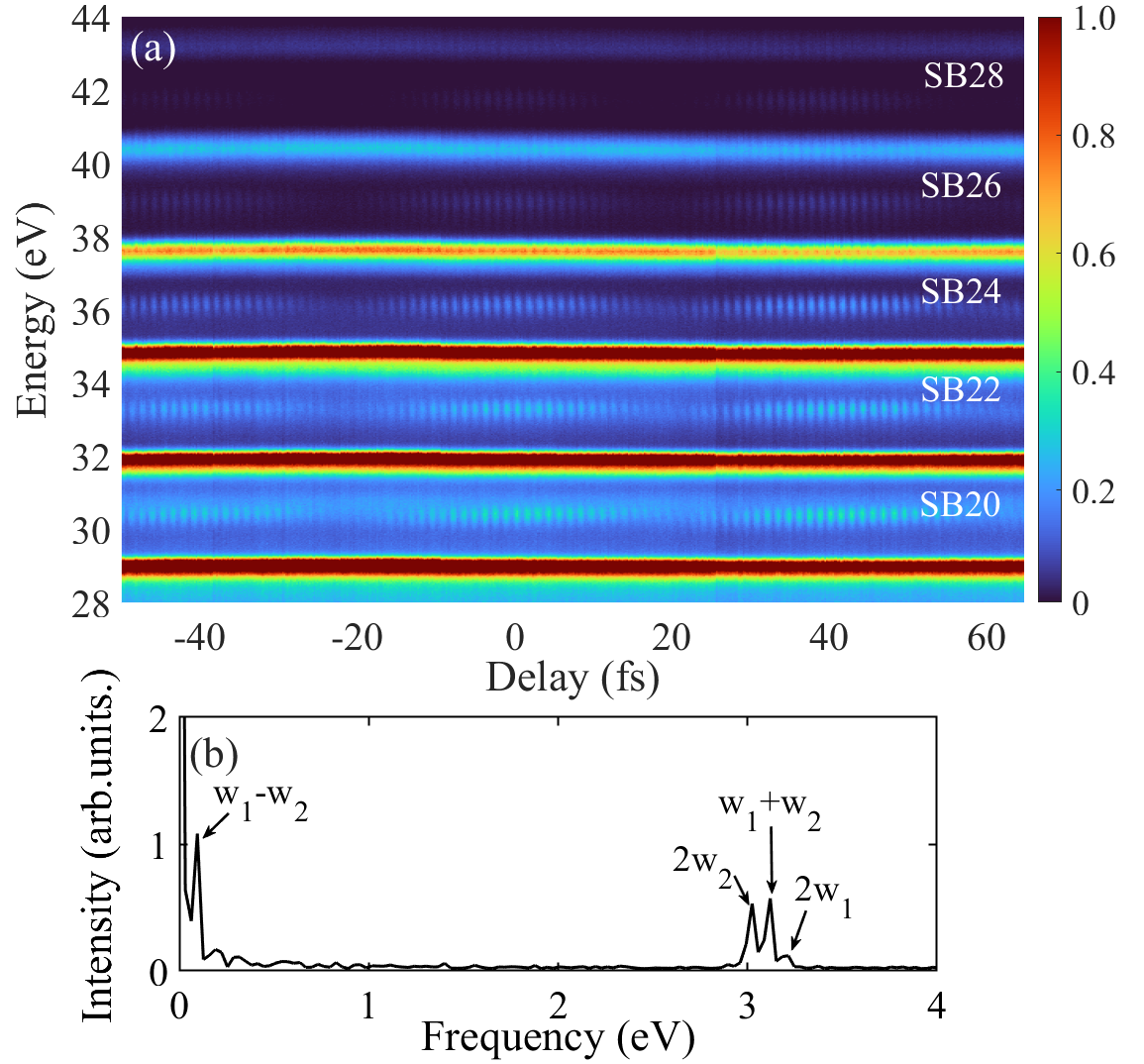}
\caption{(a) Photoelectron signal (in color) in Ar as function of harmonic photon energy and delay in a KRAKEN measurement, i.e. with a bichromatic probe ($\approx 7$\,nm bandwidths centered at 770\,nm ($\omega_1$) and 820\,nm ($\omega_2$)). The delay step is 200\,as. The total acquisition time of the scan is $\sim13$ hours. (b) Frequency components of oscillations in SB24.}
\label{figKRAKEN}
\end{figure}

Finally, our setup allows us to manipulate the pump and probe pulses independently in a simple manner. This is important for the implementation of new excitation schemes, which go beyond the interferometric RABBIT technique. One of these is the proposed KRAKEN protocol \cite{Laurell}, enabling quantum state tomography of photoelectron wavepackets, i.e. the measurement of the density matrix. This method requires bichromatic probe fields including two components with a variable frequency difference. This motivated the development of the 4f-shaper in the probe path. Fig.~\ref{figKRAKEN} presents results of a 115 fs-long scan in argon, with 200\,as delay steps. The spectrum of the bichromatic probe, shown in Fig.~\ref{figIRspectra}(b), comprises two components with $\approx 7$\,nm widths centered at 770\,nm ($\omega_1$) and 820\,nm ($\omega_2$). The observed sidebands oscillate rapidly at frequencies $2\omega_1$, $2\omega_2$ as well as $\omega_1+\omega_2$ and slowly at the beating frequency $\omega_1-\omega_2$. The amplitudes and phases of individual oscillations can be extracted via Fourier transform. The different frequency components of the SB24 oscillations are shown in Fig.~\ref{figKRAKEN}(b). 

\begin{figure}
\begin{center}
\includegraphics[width=0.5\textwidth]{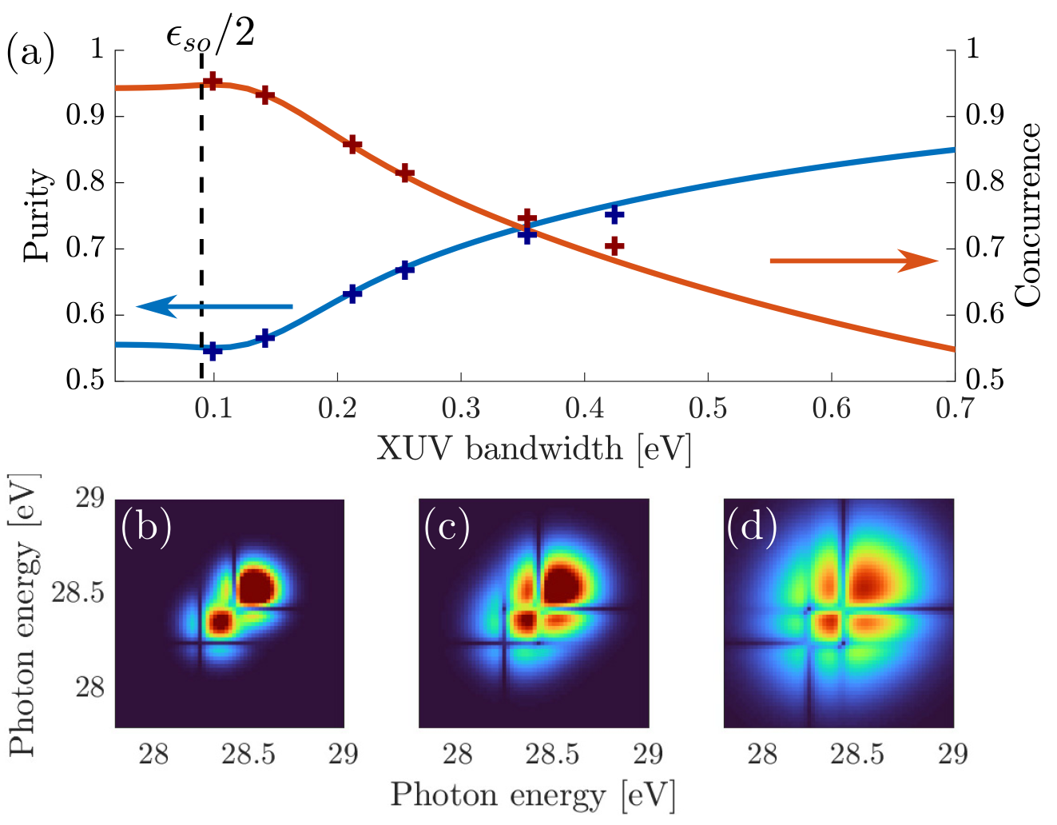}
\end{center}
\caption{Entanglement between ion and photoelectron in argon. (a) the purity and concurrence of electron from a direct calculation (blue and red lines) and KRAKEN reconstruction (blue and red crosses). (b)-(d) are the reconstructed density matrices of photoelectron from KRAKEN protocol with XUV bandwidths at 0.14 eV, 0.21 eV, and 0.35 eV \cite{Laurell}.}
\label{figKRAKENAr}
\end{figure} 

he KRAKEN protocol requires a tunable bichromatic IR probe at a variable delay relative to the XUV pulses. The density matrix of electron wavepackets can be reconstructed by performing several long delay scans (typically a few hundreds of fs) using a bichromatic IR probe with different frequency differences ($\omega_1 - \omega_2$) \cite{Laurell}. This motivated the design and commissioning of the ultrastable and flexible setup presented in this work. As indicated in the simulated results presented in Fig.~\ref{figKRAKENAr} \cite{Laurell}, both the purity of the quantum state, and the degree of electron-ion entanglement (concurrence)  can be measured by performing KRAKEN measurements. Fig.~\ref{figKRAKENAr} also shows that these quantities can be controlled in a simple way, by varying the bandwidth of the XUV radiation, which can easily be done in our setup using the spectral filter implemented in the pump path.

\section{SUMMARY}

In this article, a versatile, ultra-stable, high-spectral resolution setup for attosecond experiments based on a Mach-Zehnder interferometer is presented. An active stabilization system, which uses parts of the pump and probe femtosecond laser beams, minimizes the effects of fluctuations and drifts.
It allows for the control of the delay between the pump- and probe arms with an RMS error of 13\,as over several hours. Novel experimental schemes are enabled by independent spectral tuning of the femtosecond pulses in the pump- and probe arms, in particular allowing for bi- and polychromatic probes.  

The capability of the setup is demonstrated in three different types of experiments. The first one is a long RABBIT (interferometric) measurement showing high stability for several hours over a large delay range of 70\,fs. The second demonstrates the high spectral resolution that can be obtained by using a 10\,nm bandpass spectral filter in the probe path. A RABBIT measurement, performed in argon, with HH17 close to the 3s$^{-1}$4p autoionizing state, shows excellent spectral resolution allowing for the observation of spin-orbit splittings and Fano resonances. The analysis of the data using the Rainbow RABBIT method allows us to measure large phase jumps (close to $2\pi$) owing to the resonance.   
Finally, the flexibility of our setup is exemplified by performing a scan using a bichromatic probe field, obtained with a 4f-shaper. This will be useful for the implementation of new characterization techniques, for the complete characterization of mixed photoelectron quantum states on the attosecond time scale. In the future, full shaping capability of the probe pulses in terms of amplitude and phase can be obtained by the integration of a spatial light modulator.

\section*{Acknowledgements}

The authors acknowledge support from the Swedish Research Council (2013-8185, 2016-04907, 2018-03731, 2020-0520, 2020-03315, 2020-06384), the European Research Council (advanced grant QPAP, 884900) and the Knut and Alice Wallenberg Foundation. AL and MA are partly supported  by the Wallenberg Center for Quantum Technology (WACQT) funded by the Knut and Alice Wallenberg foundation. 

\bibliography{references.bib}

\end{document}